\documentclass[reprint,superscriptaddress,aps]{revtex4-2}

\usepackage{amsmath}
\usepackage{amssymb}
\usepackage{dcolumn}
\usepackage{graphicx}
\usepackage{xcolor}
\usepackage{graphicx}
\usepackage{bm}
\usepackage{xspace}
\usepackage{changes}
\usepackage{url}

\usepackage{hyperref}
\hypersetup{colorlinks=true, citecolor=blue, urlcolor=gray, linkcolor=cyan}
\let\oldbibitem\bibitem
\renewcommand{\bibitem}{%
  \renewcommand{\doi}[1]{\href{ http://dx.doi.org/##1}{doi:##1}}% Override \doi
  \let\bibitem\oldbibitem% Restore \bibitem
  \oldbibitem% Call old \bibitem
}

\usepackage{ulem}
\usepackage[utf8]{inputenc}   
\usepackage[T1]{fontenc}
\usepackage{lineno}
\usepackage{tabularx}

\newcommand{\CRAB}{\textsc{Crab}\xspace}

\newcommand{\IRADINA}{\textsc{Iradina}\xspace}
\newcommand{\FIFRELIN}{\textsc{Fifrelin}\xspace}
\newcommand{\FIFRADINA}{\textsc{Fifradina}\xspace}
\newcommand{\GEANT}{\textsc{Geant4}\xspace}
\newcommand{\TOUCANS}{\textsc{Toucans}\xspace}

\newcommand{\affiliationa}{\affiliation{IRFU, CEA, Universit\'e Paris-Saclay, 91191 Gif-sur-Yvette, France}}

\newcommand{\affiliationb}{\affiliation{Universit\'e Paris-Saclay, CEA, DES, SRMP, F-91191 Gif-sur-Yvette, France}}

\newcommand{\affiliationc}{\affiliation{CEA, DES, IRESNE, DER, Cadarache, F-13108 Saint-Paul-Lez-Durance, France}}

\newcommand{\affiliationd}{\affiliation{INFN, Sezione di Roma, I-00185, Roma, Italy}} %Piazzale Aldo Moro 2,

\newcommand{\affiliatione}{\affiliation{INFN, Sezione di Roma "Tor Vergata", I-00133 Roma, Italy}}

\newcommand{\affiliationf}{\affiliation{Dipartimento di Fisica, Universit\`{a} di Roma "Tor Vergata", I-00133 Roma, Italy}}

\newcommand{\affiliationg}{\affiliation{Physik-Department, Technische Universit\"{a}t M\"{u}nchen, D-85748 Garching, Germany}}

\newcommand{\affiliationh}{\affiliation{Universit\'{e} Paris-Saclay, CNRS/IN2P3, IJCLab, 91405 Orsay, France}}

\newcommand{\affiliationi}{\affiliation{TU Wien, Atominstitut, 1020 Wien, Austria}}

\newcommand{\affiliationj}{\affiliation{INFN, Laboratori Nazionali del Gran Sasso, I-67100 Assergi (AQ), Italy}}

\newcommand{\affiliationk}{\affiliation{Max-Planck-Insitut f\"{u}r Physik, D-80805 M\"{u}nchen, Germany}}

\newcommand{\affiliationl}{\affiliation{Dipartimento di Fisica, Sapienza Universit\`{a} di Roma, I-00185 Roma, Italy}}

\newcommand{\affiliationm}{\affiliation{Istituto Nazionale di Fisica Nucleare, Sezione di Roma, I-00185 Roma, Italy}}

\newcommand{\affiliationn}{\affiliation{Institut f\"ur Hochenergiephysik der \"Osterreichischen Akademie der Wissenschaften, A-1050 Wien, Austria}}

\begin{document}

%\linenumbers
\preprint{APS/123-QED}
% Main title of the paper
\title [mode = title]{Study of collision and $\gamma$-cascade times following neutron-capture processes in cryogenic detectors}                      
\author{G.~Soum-Sidikov}\affiliationa
\email{gabrielle.soum@cea.fr}

\author{H.~Abele}\affiliationi
\author{J.~Burkhart}\affiliationn
\author{F.~Cappella}\affiliationd
\author{N.~Casali}\affiliationd
\author{R.~Cerulli}\affiliatione \affiliationf
\author{A.~Chalil}\affiliationa \affiliationc 
\author{A.~Chebboubi}\affiliationc
\author{J-P.~Crocombette}\affiliationb 
\author{G.~del~Castello}\affiliationl \affiliationm
\author{M.~del~Gallo~Roccagiovine}\affiliationl \affiliationm
\author{A.~Doblhammer}\affiliationi
\author{S.~Dorer}\affiliationi
\author{E.~Dumonteil}\affiliationa
\author{A.~Erhart}\affiliationg
\author{A.~Giuliani}\affiliationh
\author{C.~Goupy}\affiliationa
\author{F.~Gunsing}\affiliationa
\author{E.~Jericha}\affiliationi
\author{M.~Kaznacheeva}\affiliationg
\author{A.~Kinast}\affiliationg
\author{H.~Kluck}\affiliationn
\author{A.~Langenkämper}\affiliationg
\author{T.~Lasserre}\affiliationa \affiliationg
\author{A.~Letourneau}\affiliationa
\author{D.~Lhuillier}\affiliationa 
\author{O.~Litaize}\affiliationc 
\author{P.~de~Marcillac}\affiliationh
\author{S.~Marnieros}\affiliationh
\author{R.~Martin}\affiliationa
\author{T.~Materna}\affiliationa
\author{E.~Mazzucato}\affiliationa
\author{C.~Nones}\affiliationa
\author{T.~Ortmann}\affiliationg
\author{L.~Pattavina}\affiliationf \affiliationj
\author{D.V.~Poda}\affiliationh
\author{L.~Peters}\affiliationg
\author{J.~Rothe}\affiliationg
\author{N.~Schermer}\affiliationg
\author{J.~Schieck}\affiliationi \affiliationn 
\author{S.~Sch\"{o}nert}\affiliationg
\author{O.~Serot}\affiliationc
\author{L.~Stodolsky} \affiliationk
\author{R.~Strauss}\affiliationg
\author{L.~Thulliez}\affiliationa
\author{M.~Vignati}\affiliationl \affiliationm
\author{M.~Vivier}\affiliationa
\author{V.~Wagner}\affiliationg
\author{A.~Wex} \affiliationg

\collaboration{\CRAB{} Collaboration}

\begin{abstract}
The emission of $\gamma$-rays after a neutron capture in a cryogenic detector can generate mono-energetic nuclear recoils in the sub-keV regime, of direct interest for the calibration of Dark Matter and Coherent Elastic Neutrino Nucleus Scattering experiments. Here we show that accurate predictions of the nuclear recoil spectra induced by neutron captures require taking into account the interplay between the development in time of the de-excitation $\gamma$-cascade of the target nucleus and that of the associated atomic collisions in matter. We present detailed simulations coupling the \FIFRELIN code for the description of the $\gamma$-cascades and the \IRADINA code for the modelling of the fast atomic movements in matter. Nuclear recoil spectra are predicted, and made available to the community, for concrete cases of  Al$_2$O$_3$, Si, Ge and CaWO$_4$ crystals exposed to a low intensity beam of thermal neutrons. We find that timing effects cause new calibration peaks to emerge in the recoil spectra and also impact the shape of the continuous recoil distribution. We discuss how they could give access to a rich physics program, spanning the accurate study of the response of cryogenic detectors in the sub-keV range, tests of solid state physics simulations and tests of nuclear models. 
\end{abstract}

\maketitle

\section{Introduction}
\label{sec:intro}

The use of nuclear recoils induced by radiative neutron capture has been recently proposed  \cite{Thulliez:2020esw} and demonstrated \cite{CRAB2022,CRESST:2023cxk} to accurately calibrate cryogenic detectors with nuclear recoils in the 100~eV -- 1~keV energy range. This technique, called \CRAB for "Calibrated Recoils for Accurate Bolometry" can be applied to most of the detectors developed in the framework of a large research program which aims to extend the search for dark matter to lighter masses, below 1 GeV/c$^{2}$ \cite{Schumann:2019eaa}, and to explore Coherent Elastic Neutrino-Nucleus Scattering (CE$\nu$NS) for original tests of the standard model at low energy \cite{Abdullah:2022zue,Lindner:2016wff,Dent:2016wcr}. To reach the desired sensitivities, macroscopic detectors of 1 to 100 grams in mass with ultra low thresholds of a few tens of eV are required \cite{CRESST:2019jnq,Strauss:2017cam,PhysRevD.99.082003,EDELWEISS:2022ktt,SuperCDMS:2020aus}. In this paper we present improved predictions of nuclear recoil spectra based on the study of fundamental timing effects of $\gamma$-emission by the excited nucleus and atomic collisions induced by its recoil. This work extends the reach of the \CRAB method to a wider energy range and to most commonly used materials in cryogenic detectors. We anticipate that this result will allow progress towards finer study of the response of cryogenic detectors but also to provide unique tests of solid state and nuclear physics.

The \CRAB calibration has the unique characteristic of using a process identical in all respects to that induced by the scattering of a dark matter particle or a neutrino, i.e. a pure nuclear recoil anywhere in the volume of the detector. The principle is very simple: a flux of thermal neutrons, typically of 0.025~eV energy, is sent onto the cryogenic detector to induce neutron capture on the nuclei of the crystal. Considering the typical orders of magnitude of the capture cross section (1 -- 10~barn) and the typical centimeter dimensions of the detectors, the capture vertices are distributed almost uniformly in the crystal. By definition a compound nucleus is formed in an excited state very close to $S_n$, the separation energy of a neutron. It then de-excites by emission of $\gamma$-rays which, by momentum conservation, makes the nucleus recoil. When the de-excitation takes place with a single $\gamma$-ray connecting directly to the ground state, this $\gamma$ easily escapes from the crystal without any interaction because of its high energy ($S_n$ values are typically between 5 and 10 MeV). Only the desired signal remains: the recoil of a nucleus, with an energy perfectly defined by the two-body kinematics
\begin{equation}
    E_{recoil} = E^2_\gamma / 2M.
    \label{eqn:recoil}
\end{equation}
As the mass $M$ and the energy $E_\gamma$ of the transition are characteristic of the emitting nucleus, each isotope entering the composition of the crystal is potentially at the origin of a calibration peak. Numerically, only a few dominant peaks are predicted for the most commonly used crystals (Si, Ge, Al$_2$O$_3$, CaWO$_4$), in an energy range of 100 to 1000~eV corresponding to that expected for light dark matter scattering or for CE$\nu$NS. Due to their very small neutron capture cross sections, oxygen and calcium isotopes have a negligible contribution to the capture rate and therefore to the recoil spectrum. The same conclusion actually applies to lead isotopes in PbWO$_4$ crystals.

The first direct detection of a \CRAB peak has been realized with CaWO$_4$ crystals and a commercial neutron source placed near the cryostat \cite{CRAB2022,CRESST:2023cxk}. The use of a purely thermal neutron beam from a research reactor as proposed in \cite{Thulliez:2020esw} will allow a drastic improvement of the signal to background ratio of these measurements. To take advantage of this potential accuracy we study here the nuclear recoils induced by multi-$\gamma$ transitions, which are in large majority compared to the single-$\gamma$ emission discussed above. Naively the nuclear recoil is then given by the vector sum of the $i$ emitted $\gamma$-ray momenta
\begin{equation}
\label{eq:vec_sum}
    E_{recoil} = \left|\vec{P_{tot}}\right|^2 / 2M,  \qquad \vec{P_{tot}}  = \sum_i \vec{P_{\gamma_i}}.
\end{equation}
This results in a continuous distribution of nuclear recoils, between some minimum value when $\gamma$-rays are emitted in opposite directions and a maximum value corresponding to the single-$\gamma$ calibration peak, theoretically reached when all $\gamma$-rays are emitted in the same direction. However, an implicit assumption of Equation~\ref{eq:vec_sum}, which we will call the "prompt" assumption, is that all $\gamma$-rays are emitted in an extremely short time compared to the duration of the nucleus recoil in matter. We will show that the time evolution of the $\gamma$ de-excitation cascades on one side and the nuclear recoil induced collisions on the other side can on the contrary be comparable, with a common order of magnitude of about 10$^{-13}$ s, and that this will have a significant impact on the recoil spectrum shape. To be convinced, let us consider the opposite extreme case where the recoiling nucleus always has time to stop before the emission of the next $\gamma$-ray in the cascade. In this "slow" case, the recoil energy induced by each $\gamma$ will be deposited sequentially so that
\begin{equation}
\label{eq:Ekin_sum}
    E_{recoil} = \sum_i E_{recoil_i} = \sum_i \left| \vec{P_{\gamma_i}} \right|^2 / 2M. 
\end{equation}
The total energy deposited is therefore no longer dependent on the direction of the emitted  $\gamma$-rays. To a given $\gamma$-cascade now corresponds a unique deposited energy, potentially offering additional calibration features in the recoil spectrum if the branching ratio of this de-excitation path is sufficiently likely.

In section~\ref{sec:fifrelin} we explain how experimental data and theoretical models are combined in the \FIFRELIN code to provide the most complete description possible of the $\gamma$-decay schemes of all isotopes of interest. In section~\ref{sec:iradina} we discuss how these $\gamma$-cascades have been interfaced with the \IRADINA code which simulates collisions and energy losses of recoiling atoms in matter. The predicted spectra are presented in section ~\ref{sec:PredRecoilSpec} and their associated physics reach discussed in section~\ref{sec:discussion}.

\section{Simulation of de-excitation $\gamma$-cascades with the \FIFRELIN code}
\label{sec:fifrelin}

\FIFRELIN is a code dedicated to the de-excitation of compound nuclei using the statistical Hauser-Feshbach framework \cite{Hauser1952} implementing the notion of nuclear realization. A nuclear realization is a set of level scheme and partial decay widths accounting for the so-called Porter-Thomas fluctuations of reduced transition probabilities~\cite{Porter1956}. Originally it has been developed to simulate and understand the nuclear fission process \textit{via} the de-excitation of the fission fragments \cite{Litaize2015}. Recently its accurate predictions of $\gamma$ and electron spectra after radiative thermal neutron captures have been successfully applied to neutrino physics \cite{Almazan2019,Almazan:2022tzc}. 

\begin{figure}[h]
	\centering
	\includegraphics[width=1.0\linewidth]{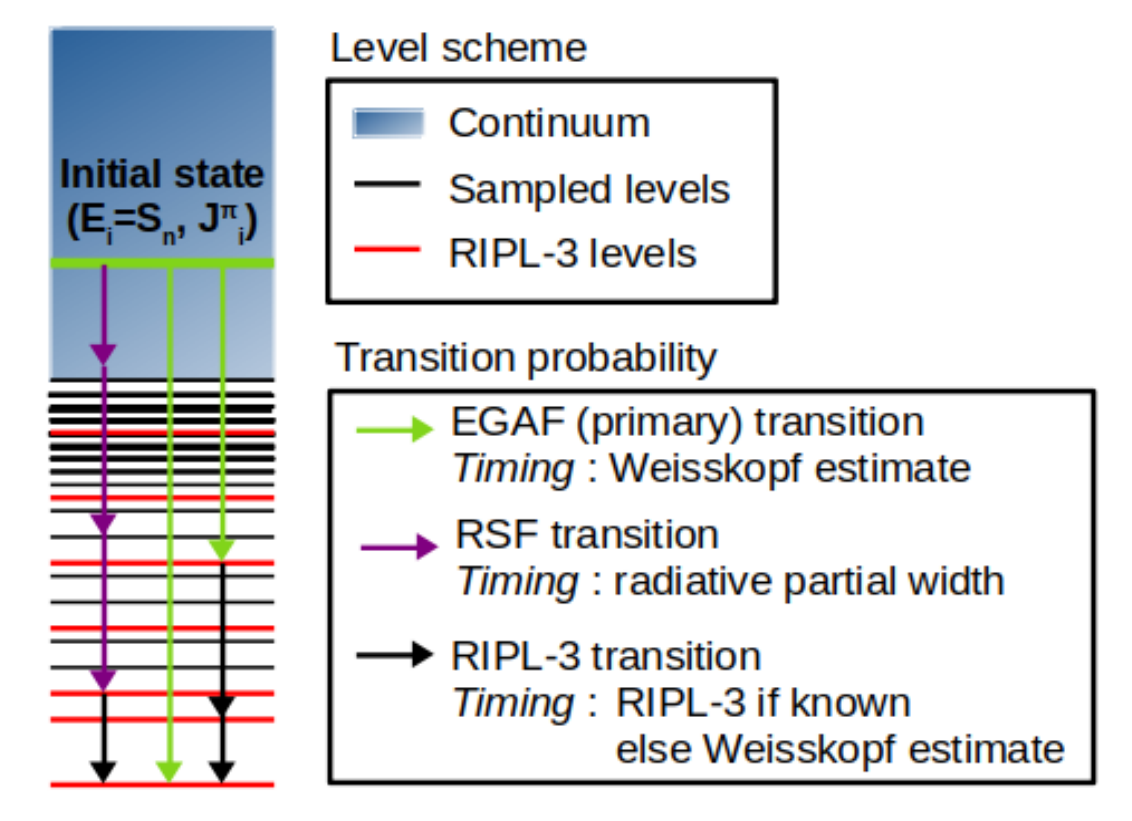}
	\caption{Illustration of the different nuclear inputs used by FIFRELIN to build and simulate the de-excitation of a nucleus.}
	\label{fig:fifrelin_decay_scheme}
\end{figure}
 
The \FIFRELIN code algorithm, whose principle is sketched in Figure~\ref{fig:fifrelin_decay_scheme}, can be divided in three steps. Firstly, the compound nucleus initial state is defined by the nuclear selection rules based on the target and incoming neutron properties. Secondly, a nuclear level scheme of the nucleus is built (one nuclear realization). All the available data at low energies are taken from the \mbox{RIPL-3} database \cite{RIPL3} and those related to the initial state (primary $\gamma$'s) are taken from the EGAF database \cite{EGAFpubli}. For light nuclei such as aluminium or silicone isotopes, the nuclear level scheme, the transition probabilities and most of the level half-lives are in the databases, therefore the computation of a $\gamma$ cascade is rather straightforward. However this is not as easy for medium and heavy nuclei such as germanium or tungsten isotopes where the number of nuclear levels is too high and cannot be experimentally determined. Therefore nuclear level density models are used to complete the level scheme. Thirdly, the de-excitation of the nucleus is performed based on the knowledge of the transition probability $P(\gamma_{i \rightarrow f})$ to go from an initial level $i$ to a final level $f$. A transition is completely characterized by its type X~=~{E, M}, multipolarity L~=~{1, 2, 3...} and energy $E_{\gamma}$. The energy, spin and parity are therefore conserved on a step-by-step basis. If the transition probability is not available in any database, it can be computed from the partial width $\Gamma_{\gamma_{i \rightarrow f}}$ according to
\begin{equation}
P(\gamma_{i \rightarrow f}) = \frac{\Gamma^{RSF}_{\gamma_{i \rightarrow f}}}{\sum_{f'} \Gamma^{RSF}_{\gamma_{i \rightarrow f'}}},
\end{equation}
and
\begin{equation}
\label{eq:RSF_formula}
      \Gamma^{RSF}_{\gamma_{i \rightarrow f}} = \langle\Gamma^{RSF}_{\gamma}\rangle\, \epsilon_{PT} = \overleftarrow{f}_{XL}(E_{\gamma})\, \frac{E_{\gamma}^{2L+1}}{\rho_i} \, \epsilon_{PT}.
\end{equation} 
The initial nuclear level density $\rho_i$ and the radiative strength function $\overleftarrow{f}_{XL}(E_{\gamma})$ (RSF) allow to compute an average partial width $\langle \Gamma^{RSF}_{\gamma} \rangle$ to which a Porter-Thomas fluctuation $\epsilon_{PT}$ \cite{Porter1956} is added to get $\Gamma^{RSF}_{\gamma_{i \rightarrow f}}$. 
To take into account nuclear model uncertainties, \FIFRELIN performs this algorithm multiple times, \textit{i.e.} on different nuclear realizations. More details about the algorithm can be found in \cite{Regnier2016,Almazan2019}. By default \FIFRELIN uses the Composite Gilbert-Cameron Model (CGCM) of nuclear levels density to build the level scheme, the Enhanced Generalized Lorentzian (EGLO) RSF \cite{RIPL3} to compute $\gamma$ emission probabilities and the BrIcc code to generate coefficients of $\gamma$ to electron internal conversion~\cite{Kibedi2008}. As a complementary approach to these phenomenological models microscopic models can also be considered: Hartree-Fock-Bogoliubov (HFB) calculations implementing the BSK14 Skyrme effective interaction~\cite{Goriely2008} for nuclear levels density and Quasi-particle Random Phase Approximation (QRPA) calculations for RSF (see section \ref{sec:discussion}). Further details about these models can be found in \cite{RIPL3} and references therein.

\begin{figure}
	\centering
	\includegraphics[width=1.0\linewidth]{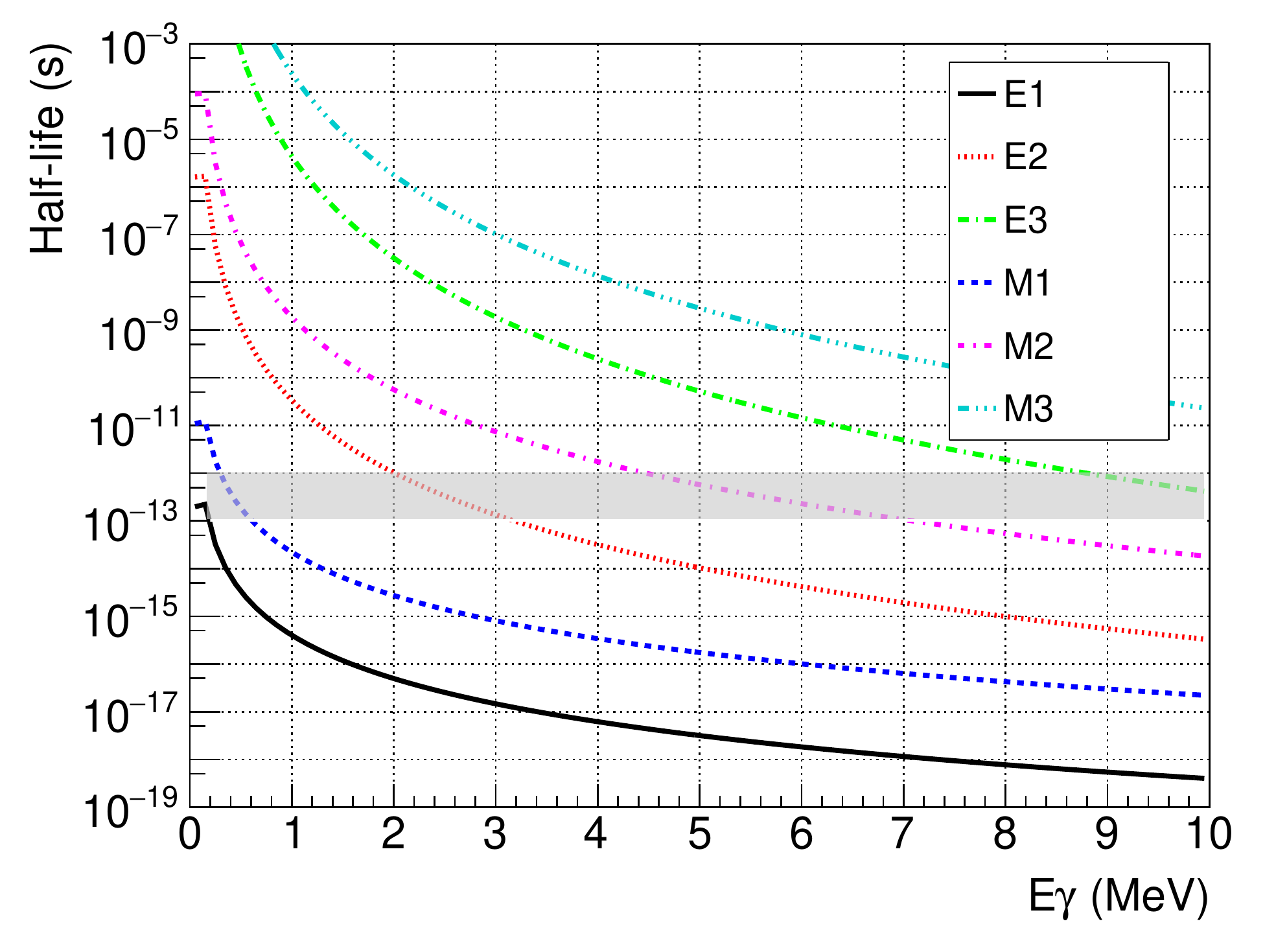}
	\caption{Application of the Weisskopf formula \cite{Preston1962} to the half-life of $\gamma$-transitions of $^{71}$Ge for various types and multipolarities. The gray shaded area indicates the typical range of stopping times of a Ge atom recoiling with 400~eV initial energy.}
	\label{fig:Weisskopf}
\end{figure}

\begin{figure}
	\centering
	\includegraphics[width=1.0\linewidth]{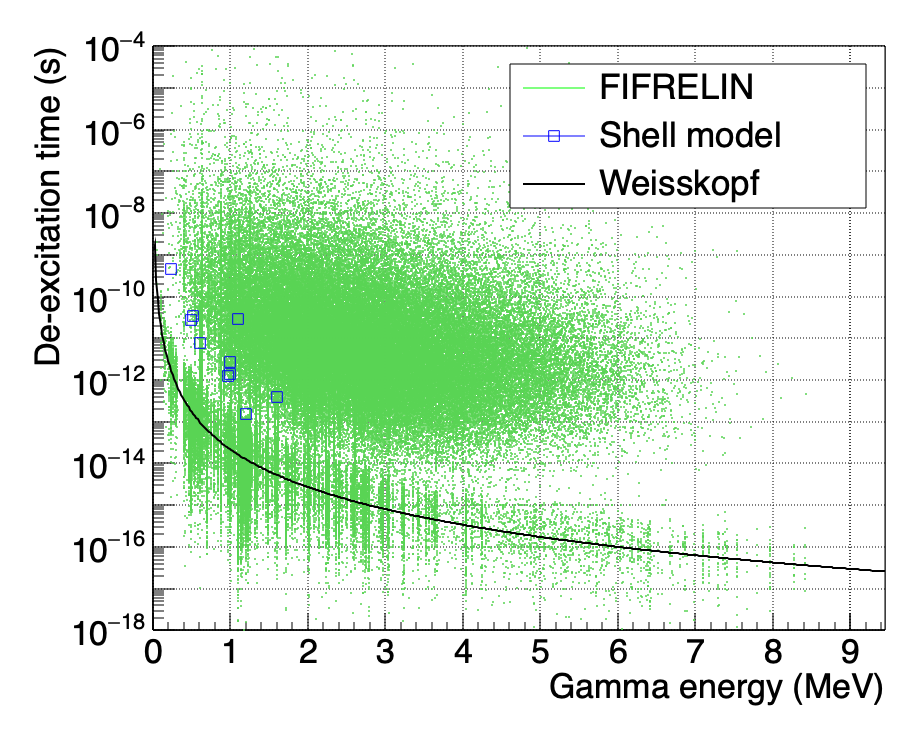}
	\caption{Comparison of the de-excitation time from \FIFRELIN (green points), the nuclear level half-life from Weisskopf estimate (black curve) and the shel model calculations (blue points) for $^{74}$Ge M1 transitions. For a given transition, the \FIFRELIN de-excitation times are drawn in an exponential distribution leading to a vertical distribution of points. The distribution of \FIFRELIN times computed with the Weisskopf estimate is centered around the black curve as expected and systematically lower than the distribution of times based on the EGLO-RSF approach.}
	\label{fig:EGLO_vs_Weisskopf_74Ge}
\end{figure}

In this work we are interested in predicting the time sequence of the de-excitation cascades. The nuclear level half-lives are primarily taken from the \mbox{RIPL-3} database~\cite{RIPL3}. If no experimental value is available, a half-life is computed from the radiative partial width of the transition using the Heisenberg uncertainty principle. Two cases arise. If the nuclear level branching ratios are known, \FIFRELIN does not compute the transition probabilities $\Gamma^{RSF}_{\gamma_{i \rightarrow f}}$, but instead it computes the Weisskopf single-particle estimates of the partial width $\Gamma^{W}_{\gamma_{i \rightarrow f}}$. In this approximation a single nucleon transition is responsible for the radiation emission leading to simple formula of $\Gamma^{W}_{\gamma_{i \rightarrow f}}$, which depends only on XL and $E_\gamma$ \cite{Preston1962}. Application of these formula to the isotope $^{71}$Ge is presented in Figure~\ref{fig:Weisskopf} for various types of transitions.
However, the simple form of $\Gamma^{W}_{\gamma_{i \rightarrow f}}$ implies neglecting nuclear collective effects that has a large impact on the transition probability and, consequently, on the nuclear level half-lives. This was already underlined in 1975 by Jones and Kraner~\cite{Jones1975} when discussing the recoil energy spectrum of germanium isotopes after a neutron capture. For excited levels with no known half-life nor branching ratio, the \FIFRELIN code calculates the transition probabilities (\textit{i.e.} $\Gamma^{RSF}_{\gamma_{i \rightarrow f}}$) using RSF models in which the collective effects are accounted for in an effective (microscopic) way as the RSF model parameters are fitted to a large set of Giant Dipole Resonance data~\cite{RIPL3}. The impact of this improved treatment of the nuclear level half-lives by \FIFRELIN is illustrated in Figure~\ref{fig:EGLO_vs_Weisskopf_74Ge} for the $^{74}$Ge de-excitation with M1 transitions. It shows that taking into account the collective effects in an effective way leads to an increase of the predicted half-lives by several orders of magnitude with respect to the Weisskopf estimate. In principle this RSF-based approach is relevant in the statistical energy domain, \textit{i.e.} at high energy, and could fail at low energy. As a complementary approach shell model calculations have also been performed with the KShell code \cite{SHIMIZU2019372} and the JUN45 effective interaction \cite{PhysRevC.80.064323}. The large computational time restricts the application of this method to few low energy levels above the ground state and also the predicted level energies have limited accuracy. Still the results of these independent calculations are found somewhat in between the Weisskopf and the EGLO-RSF distributions. We recover the known fact that the Weisskopf estimate is underestimating the half-lives for most transitions, low energy E2 transitions being the main exception \cite{Dommelen2012QuantumMF}.

\section{Implementation of in-flight gamma emission in the \IRADINA code}
\label{sec:iradina}

With the nuclear level half-lives in hand, the in-flight emission of $\gamma$-rays by a recoiling nucleus can be simulated. The simple Binary Collision Approximation (BCA) approach has been chosen since it allows for a very quick modelling of fast atomic movements in a material. It relies on the assumption that atomic trajectories can be divided in series of two body collissions, with moving atoms having straight and constant velocity trajectories between collisions. Atomic collisions are dealt with through the short range universal Ziegler potential~\cite{RN913}. The BCA approximation is valid only in the so-called ballistic phase, during which some atoms move much faster than the average speed of regularly vibrating atoms at the temperature of the material. It cannot describe the subsequent thermal phase which accounts for the collective atomic movements after the ballistic phase. While BCA is known to poorly reproduce the actual number of displaced atoms and created defects, it proves to be an efficient way to estimate time of flights of fast moving particles. 
Among the various BCA codes available, \IRADINA retained our attention as it is an open-source code written in C, making it easy to modify to suit our needs. \IRADINA has been developed by C. Borschel and C. Ronning to simulate ion beam irradiation of nanostructures \cite{Borschel2011}. 

In \IRADINA, the distribution of atoms is isotropic and homogeneous, ignoring the crystal structure of the material. Each projectile is transported in matter from collision to collision recursively until its kinetic energy falls below a given threshold, beyond which it is assumed to deposit its remaining energy and stop. Each collision step starts by randomly drawing a flight length from a Poisson distribution centered at the mean inter-atomic distance in the material. The flight-path is considered straight between successive collisions and an electronic energy loss is subtracted from the kinetic energy (see below). Collisions happen elastically between the projectile and one target atom. In the case of a compound material, the target element is chosen randomly according to the stoichiometry. An impact parameter is first randomly drawn from a uniform distribution between 0 and $b_{max}$, with $b_{max}$ such as 
\begin{equation}
\label{eq:impact_param}
N \times \pi l b_{max}^2 = 1,
\end{equation}
where $l$ is the flight-path, and $N$ is the density of the material, meaning that on average one collision target is included in the cylinder of height $l$ and radius $b_{max}$  \cite{schiettekatte2016}. The scattering angle is obtained from prepared tables to optimize the computing time, as in \textsc{Corteo} \cite{schiettekatte2008}. Finally the azimuth angle is randomly chosen from a uniform distribution between $0$ and $2\pi$. The projectile direction --- represented by the unit velocity --- is then rotated along the chosen scattering and azimuth angles and the energy transferred to the target is deduced from momentum and energy conservation. This so-called "nuclear energy loss" is subtracted from the projectile kinetic energy. If the energy transferred to the collision partner is greater than its displacement threshold defined in the input file of the simulation, it becomes a secondary projectile, thus increasing the number of displaced atoms and leaving a vacancy behind. It is further transported by the same function called recursively. Regarding the initial projectile, depending on its remaining energy, it can either stop at the collision site or move on to the next collision, until its kinetic energy becomes lower than the stopping energy.

During the flight of moving particles, the electronic energy losses are accounted for by reducing the kinetic energy of the moving particle between two subsequent collisions. The decrease of the energy is simply obtained by multiplying the flight path with the electronic stopping power, which depends on  the composition and density of the material. It is calculated from tabulated data extracted from Stopping Ranges of Ions in Matter (SRIM) databases. \cite{RN1565}.

For \FIFRELIN\!-\,\IRADINA simulations, the initial projectile is a nucleus which recoils in the detector material under the impulse given by the sequential emission of  $\gamma$-rays. In \IRADINA, the target material is described by its density and chemical composition only, with no isotopic information. However, successive simulations can be run for each projectile isotope, as \FIFRELIN provides a list of possible $\gamma$ cascades for each capturing isotope, with the number of emitted $\gamma$, their energies, and the emission timing. Each CRAB projectile starts recoiling with a kinetic energy corresponding to the emission of the first $\gamma$ of the cascade as defined by Equation~\ref{eqn:recoil} and its recoiling direction is read from the user-written configuration file. For our calculations, it can always be set to $\left(1, 0, 0\right)$ since the material is considered as isotropic. The emission of the first $\gamma$ also defines the origin of times.

For each collision step of the initial projectile, after \IRADINA selection of the flight length, a flight duration is computed as the ratio of the flight distance to the velocity, extracted from the kinetic energy. Added to the current time of flight, it is compared to the $\gamma$ emission times of the associated \FIFRELIN cascade. Any $\gamma$ emitted during the current flight is actually considered as emitted at the beginning of the \IRADINA step. Thus the energy and momentum of the projectile are corrected for by the addition of the momentum induced by the $\gamma$ emission. The polar and azimuth angles of the $\gamma$, $\theta$ and $\phi$, are randomly drawn from uniform distributions, between $\left[-1, 1\right]$ for $cos\left(\theta\right)$ and $\left[0, 2\pi\right]$ for $\phi$. Then the flight and the collision proceed as they normally do in \IRADINA. 

In case the initial projectile stops before the $\gamma$-cascade was finished, the current time is set to the emission time of the next $\gamma$, and the energy and direction of the de-exciting nucleus are reinitialized using momentum conservation. The transport function with the collision loop is then called recursively for the new recoil. 

Finally, the coupling of \FIFRELIN data and \IRADINA (called \FIFRADINA in the following) provides, for each $\gamma$-cascade, a deposited energy from nuclear recoil(s), taking both collision and cascade timings into account. Merging this information with \FIFRELIN predicted cascades, we build output files containing, for each cascade, the number of emitted $\gamma$, their energies, emission timings and angles, the number of emitted conversion electrons with their energies and emission timings, as well as the total nuclear recoil energy expected for the cascade. These files are made available in the supplemental material \cite{FifradinaData} for the isotopes discussed in section \ref{sec:PredRecoilSpec}.

However, for any practical application, the energy deposited in the cryogenic detector may contain more than the pure nuclear recoil energy. In particular low energy $\gamma$'s or conversion electrons involved in multi-$\gamma$ de-excitations might interact inside the detector and saturate the signal with energy depositions of the order of keV. By definition this so-called "internal background" has no direct impact on single-$\gamma$ calibration peaks, but it has to be taken into account for describing the global energy spectrum and counting rate expected in a detector with a given geometry. 

For our study, this is done with the \CRAB-\GEANT simulation package \cite{Thulliez:2020esw} inherited from the \TOUCANS code~\cite{THULLIEZ2023}. The standard \GEANT treatment of a nucleus de-excitation is replaced by the \FIFRADINA (or \FIFRELIN) predictions using a dedicated library made available online on the following GitLab repository~\cite{Fifrelin4Geant4}. For each neutron capture in the detector simulated by \GEANT, a cascade is read from the corresponding \FIFRADINA data file and all predicted $\gamma$-rays, conversion electrons and total nuclear recoil are propagated from the capture vertex. 
\newline
\indent
To cross-check the implementation of the \FIFRELIN $\gamma$~timings in \IRADINA, we have generated two mock \FIFRELIN timing data: in a first dataset all emission times are defined as 0~s to emulate the prompt hypothesis, while a second dataset with a fixed 1~s delay between two $\gamma$ emissions emulates the slow hypothesis. \FIFRADINA simulations on these datasets, followed \CRAB-\GEANT simulations provide a spectrum of deposited energy for each hypothesis. These two extreme cases can also be studied with our previous simulation package used in \cite{Thulliez:2020esw}, where the total nuclear recoil is provided by Equation~\ref{eq:vec_sum} for the prompt hypothesis or Equation~\ref{eq:Ekin_sum} for the slow hypothesis. Exact same spectra are obtained from these two independent approaches for the three detectors on which they have been tested (Si, Ge and CaWO$_4$), validating the new simulation framework. 

\section{Predicted nuclear recoil spectra}
\label{sec:PredRecoilSpec}
\begin{figure}
	\centering
	\includegraphics[width=1.0\linewidth]{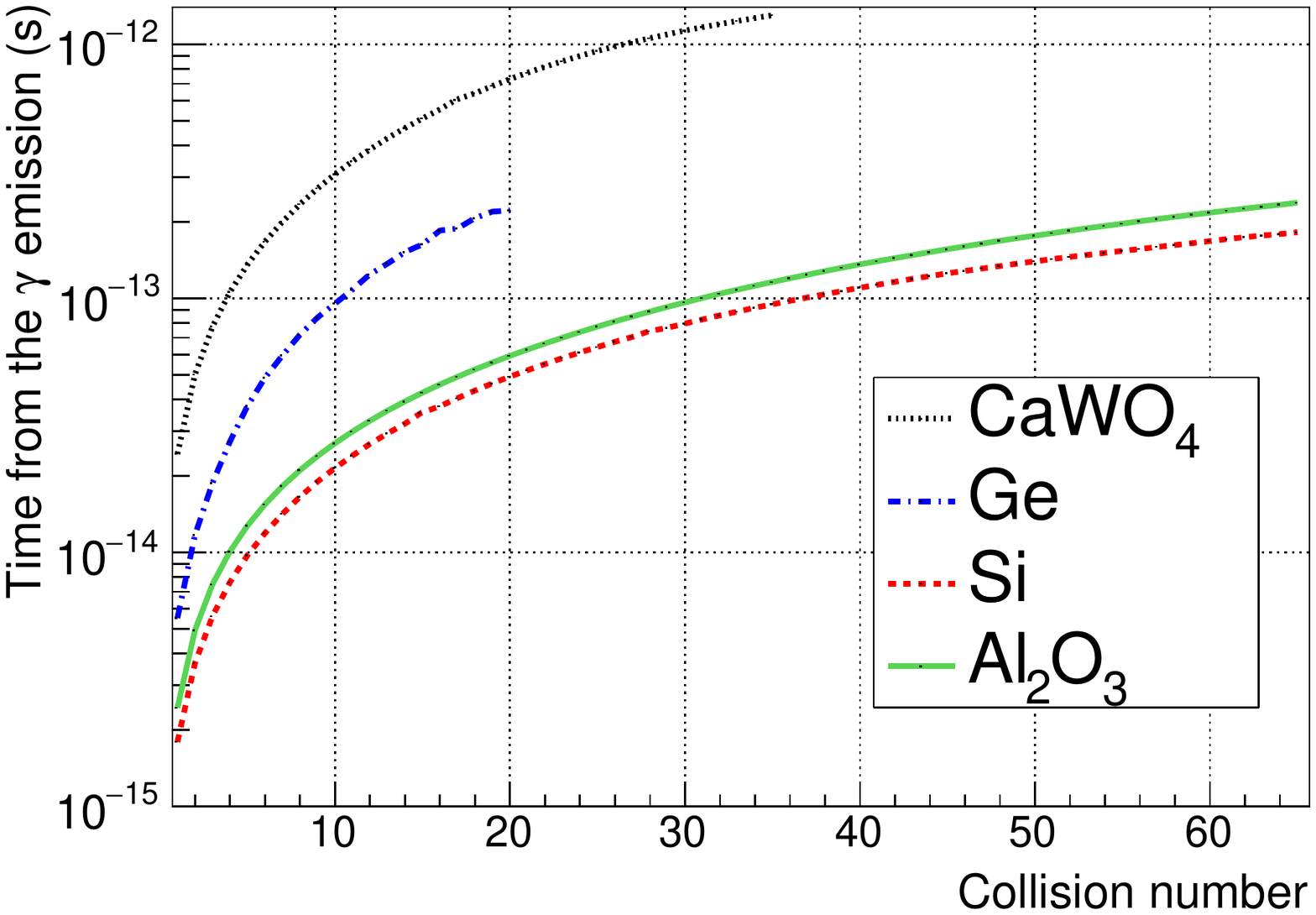}
	\caption{\IRADINA prediction of the mean time elapsed since the emission of the first $\gamma$ shown as a function of the number of collisions for various recoils induced by single-$\gamma$ transitions: 112~eV recoil of a W atom in CaWO$_4$ (dotted black), 416~eV recoil of a Ge atom in pure Ge (dashed dotted blue), 1330~eV recoil of a Si atom in pure Si (dashed red) and 1145~eV recoil of an Al atom in Al$_2$O$_3$(plain green). The end point of each simulation corresponds to the number of collisions after which 90\% of the initial kinetic energy has been lost. See Table~\ref{tab:calib_cascades} for more details on the origin of the recoil energies considered above.}
	\label{fig:Slowdown}
\end{figure}

In the following we present the results obtained for Al$_2$O$_3$, Si, Ge and CaWO$_4$ which cover a large majority of the cryogenic detectors currently used by the community. Figure~\ref{fig:Slowdown} illustrates the slowing down process of the target nuclei for typical recoil energies induced by radiative thermal neutron capture in each material. Some orders of magnitude of interest appear: the time of flight before the first collision is about 10$^{-14}$ s and the time to lose 90\% of the initial energy is between few 10$^{-13}$ s and 10$^{-12}$ s. 

Comparing the typical duration of a collision series (10$^{-14}$ to 10$^{-12}$ s) with the half-life estimates of $\gamma$-transitions discussed in section \ref{sec:fifrelin}, we see that the shift from a recoil energy given by Equation~\ref{eq:vec_sum} (prompt hypothesis) to that given by Equation~\ref{eq:Ekin_sum} (slow hypothesis) can occur when low-energy transitions are involved even for the most probable low-multipolarities. A new structure in the observed recoil spectrum will thus potentially be visible for the most likely cascades combining an energetic $\gamma$-ray with one or more low-energy $\gamma$'s to complete the total energy at $S_n$. From this criterion, we have identified in our simulations a short list of candidate $\gamma$-cascades, shown in Table~\ref{tab:calib_cascades}. All cascades in this table are expected to induce visible mono-energetic nuclear recoils either because only one $\gamma$-ray is emitted or because all $\gamma$'s following the primary $\gamma$ are emitted after the nucleus has stopped in the crystal due to the long enough half-life of the corresponding nuclear level. For most of these multi-$\gamma$ cascades the half-lives are actually evaluated, \textit{i.e.} based on experimental data. For the other half-lives the Weisskopf estimate is used. Since this approximation understimates the half-lives, we consider that it provides conservative results regarding the peak features to be expected in the nuclear recoil spectra.

\begin{table*}
\caption{Main parameters of few decay cascades identified as good candidates to induce prominent calibration features in the nuclear recoil spectra. The first two columns indicate the composition and the size of the simulated cubic crystals. The target isotope in the 3$^{rd}$ column is by definition before the neutron capture, therefore all subsequent decay data correspond to the isotope with one more neutron. The Figure Of Merit (F.O.M.) in the 4$^{th}$ column is defined as the product of the natural abundance of the target isotope, its cross section for capture of thermal neutrons and the intensity of the decay cascade considered, as proposed in \cite{Thulliez:2020esw}. Thus it is proportional to the expected number of counts in the associated recoil peak. Column 5 and 6 show the energy and half-life of all $\gamma$-transitions respectively. The half-life of the first $\gamma$-ray in a cascade is not relevant here as the emission time, whatever it is, simply defines the start time of the collision series. A measured half-life is shown with its uncertainty (or upper limit) and taken from the \mbox{RIPL-3} database \cite{RIPL3}, the notation "W" means that the Weisskopf estimate is used when no evaluate data is available. The subsequent, non-primary, $\gamma$'s are shown with a "$\hookrightarrow$" symbol in front of their energy. For each crystal type the $\gamma$-cascades are displayed in increasing order of the induced nuclear recoil (column 7).} \label{tab:calib_cascades}

\renewcommand*{\arraystretch}{1.3}
\begin{ruledtabular}
\begin{tabular}{crrrrrr}
Detector & Size & Target & F.O.M. & E$_\gamma$ & Half-life & Nuclear\\
Crystal & (mm$^{3}$) & Isotope &  & (keV) & (ps) & Recoil (eV)\\
\hline
Al$_2$O$_3$ & $5\times 5\times 5$ & $^{27}$Al & 79 & 7693 & - & \\
   &                        &           &     & $\hookrightarrow$ 30.6 & 2070 $\pm$ 50 & 1135.7 \\

   &                        &           & 616 & 7724 & - & 1144.8 \\
\hline
Si & $20\times 20\times 20$ & $^{28}$Si & 116 & 7200 & - & \\
   &                        &           &     & $\hookrightarrow$ 1273 & 0.29 $\pm$ 0.01 & 990.4 \\

   &                        & $^{28}$Si & 36 & 8474 & - & 1330.1 \\
\hline
Ge & $20\times 20\times 20$ & $^{74}$Ge & 220 & 6253 & - & \\
   &                        &           &    & $\hookrightarrow$ 253 & 1.36 (W) & 280.6 \\
   &                        & $^{70}$Ge & 261 & 6117 & - & \\
   &                        &           &    & $\hookrightarrow$ 1299 & 0.42 $\pm$ 0.09 & 296.0 \\
   &                        & $^{74}$Ge &  54 & 6506 & - & 303.2 \\
   &                        & $^{70}$Ge & 287 & 6708 & - & \\
   &                        &           &    & $\hookrightarrow$ 708 & <10.70 & 344.3 \\
   &                        & $^{70}$Ge & 238 & 6916 & - & \\
   &                        &           &    & $\hookrightarrow$ 500 & 0.18 (W) & 363.9 \\
   &                        & $^{70}$Ge & 122 & 7416 & - & 416.2 \\
   &                        & $^{73}$Ge &  117 & 8733 & - & \\
   &                        &           &     & $\hookrightarrow$ 868 & 1.53 $\pm$ 0.10 & \\
   &                        &           &     & $\hookrightarrow$ 596 & 12.41 $\pm$ 0.09 & 561.8 \\
\hline
CaWO$_4$ & $5\times 5\times 5$  & $^{186}$W &  2260 & 5262 & - &  \\
        &                      &           &    & $\hookrightarrow$ 205 & 2.6 (W) & 79.6 \\

        &                      & $^{186}$W &  2159 & 5321 & - &  \\
        &                      &           &    & $\hookrightarrow$ 146 & 7.1 (W) & 81.4 \\
         &                     & $^{182}$W &  7506 & 6191 & - & 112.5 \\
         &                     & $^{183}$W &  823 & 7411 & - & 160.3 \\

\end{tabular}
\end{ruledtabular}
\end{table*}

In the following we discuss the practical cases of cryogenic detectors with dimensions shown in Table~\ref{tab:calib_cascades}. The spectra of deposited energy after \GEANT simulation are provided. The critical impact of energy resolution is discussed in the last section. The simulated neutron beam is always purely thermal neutrons, perpendicular to one face of the crystal cube and with a circular section inscribed in the detector section. 

\begin{figure}
	\centering
	\includegraphics[width=1.0\linewidth]{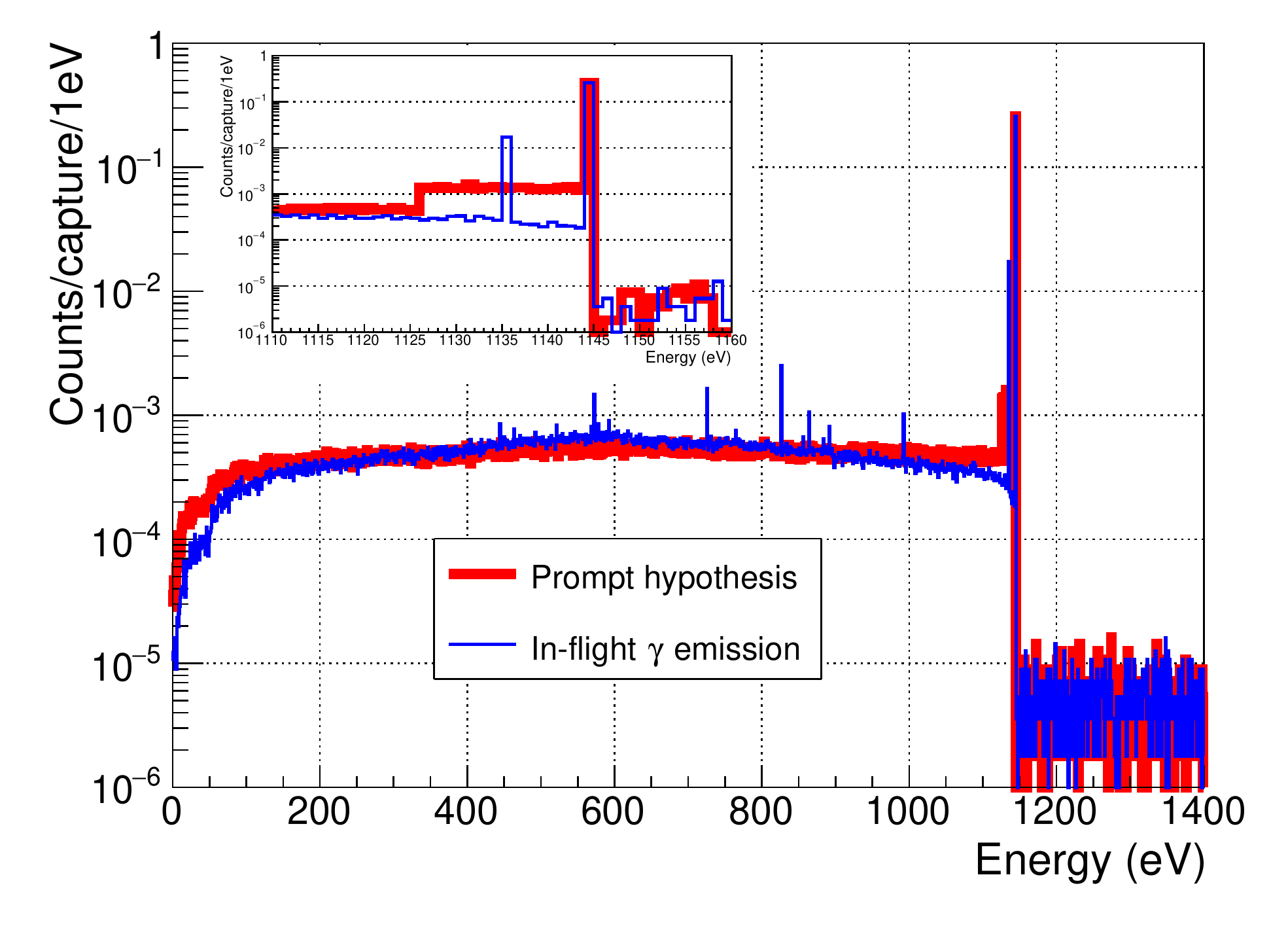}%{FIGS/Fig_Al_Norm.pdf}
	\caption{Predicted nuclear recoil spectra in a 5~$\times$~5~$\times$~5~mm$^3$ Al$_2$O$_3$ crystal using the prompt hypothesis (red) or our best estimate of timing effects from the \FIFRADINA-\GEANT software (blue). The inset shows a zoom around the main recoil line at 1145~eV. The events on the right of the single-$\gamma$ peak are due to $\gamma$'s or conversion electrons interacting in the crystal and depositing energy on top of the nuclear recoil. This simulation was obtained with 10$^8$ incident neutrons corresponding to 5.575$\times$10$^5$ neutron captures in the crystal. The $\gamma$ emission times are sampled by \FIFRELIN using the half-life times from the RIPL-3 database.}
	\label{fig:Al_recoil_pred}
\end{figure}

Starting with the lightest target isotope $^{27}$Al, Figure~\ref{fig:Al_recoil_pred} shows the predicted nuclear recoil spectrum expected in an Al$_2$O$_3$ crystal exposed to a flux of thermal neutrons. For a light isotope such as $^{28}$Al all excited levels are known and indexed in the nuclear databases with their half-lives. The 1145~eV recoil peak induced by the single-$\gamma$ transition is quite intense due to the unusually large branching ratio of this transition (26.8\%), compensating for the small neutron capture cross section. The timing effects discussed above make a few more recoil peaks appear above the flat continuous distribution of the multi-$\gamma$ cascades. However their small intensities will make them be rapidly smeared out by the resolution effects. An interesting textbook case is illustrated in the inset of Figure~\ref{fig:Al_recoil_pred}. In the prompt hypothesis a step on the left of the single-$\gamma$ peak is visible due to a two-$\gamma$ cascade with the energy of the primary $\gamma$ quite close to S$_n$. When the emission time of the second $\gamma$ is taken into account, all events in this step are stacked in one peak with an energy of 1136~eV defined by Equation~\ref{eq:Ekin_sum}, see Table~\ref{tab:calib_cascades}. 

\begin{figure}
	\centering
	\includegraphics[width=1.0\linewidth]{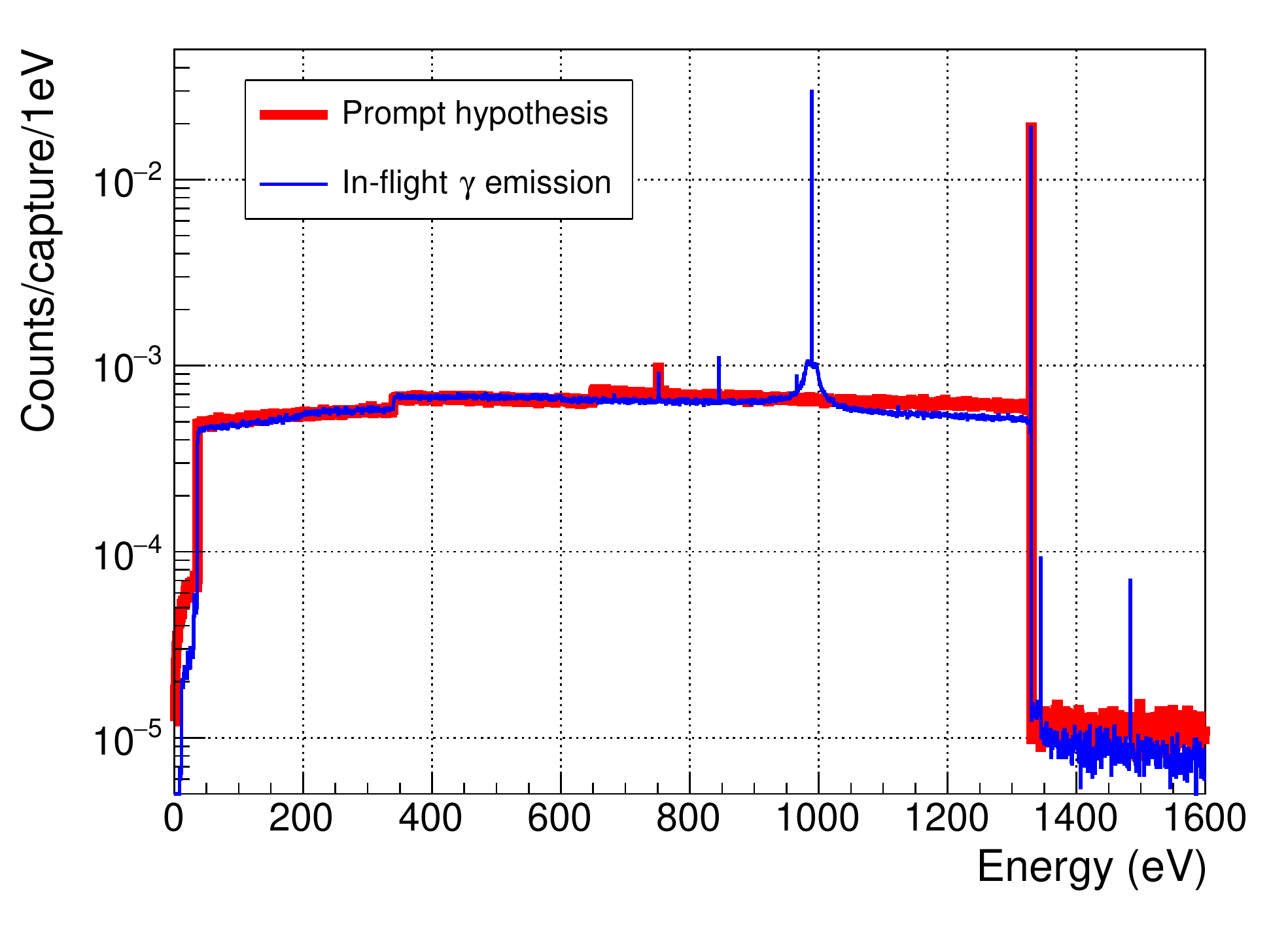}
	\caption{Predicted nuclear recoil spectra in a 20~$\times$~20~$\times$~20~mm$^3$ Si crystal using the prompt hypothesis (red) or our best estimate of timing effects from the \FIFRADINA-\GEANT software (blue). The timing effects induce a prominent peak at 990~eV with a Dirac-like structure on top of a broader pedestal. The events on the right of the single-$\gamma$ peak at 1330~eV are due to neutron captures on the other silicon isotopes, with significantly smaller natural abundance. This simulation was obtained with 4$\times$10$^8$ incident neutrons corresponding to 6.506$\times$10$^6$ neutron captures in the crystal. The $\gamma$ emission times are sampled by \FIFRELIN using the half-life times from the RIPL-3 database.}
	\label{fig:Si_recoil_pred}
\end{figure}

A similar configuration is predicted for $^{28}$Si, the most abundant isotope of natural silicon, with a recoil peak at 1330~eV associated to the single-$\gamma$ transitions (see Figure~\ref{fig:Si_recoil_pred}). In this case, the timing effects induce an extra prominent peak centered at 990~eV. It corresponds to a 2-$\gamma$ cascade where the half-life of the intermediate excited level turns out to be very similar to the stopping time of the Si atom recoiling after the first $\gamma$ emission. This enriches the structure of the associated calibration peak since the second $\gamma$ can be emitted either in flight or after the stop of the Si atom with comparable probabilities. Thus part of the events accumulates in the mono-energetic peak as predicted by Equation~\ref{eq:Ekin_sum} and another part forms a broader distribution, like a pedestal to the peak, due to the fact that the second $\gamma$ boosts or decelerates the recoiling Si atom depending on their relative direction of motion. This fine-tuning of the timing of the $\gamma$ and collision cascades is quite unique and an independent study leads to a similar prediction \cite{Harris:2023tdp}. In the next section we discuss how this feature could be used as a sensitive probe of the underlying solid state physics. 

\begin{figure}
	\centering
	\includegraphics[width=1.0\linewidth]{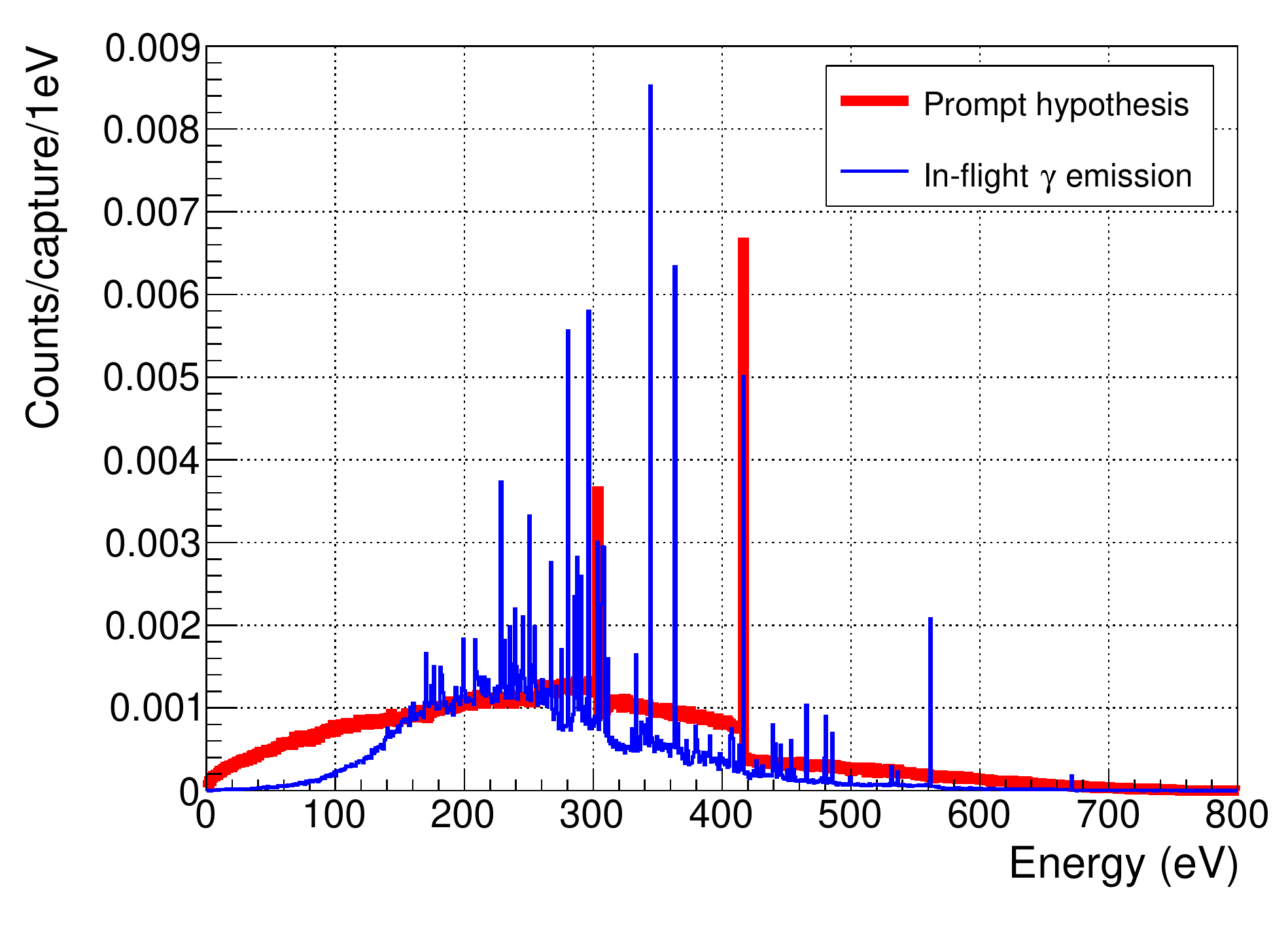}
	\caption{Predicted nuclear recoil spectra in a 20~$\times$~20~$\times$~20~mm$^3$ Ge crystal using the prompt hypothesis (red) or our best estimate of timing effects from the \FIFRADINA-\GEANT software (blue). The timing effects reveal a lot richer structure of the recoil spectrum exploitable for calibration at low energy. This simulation was obtained with 10$^7$ incident neutrons corresponding to 1.86$\times$10$^6$ neutron captures in the crystal. The $\gamma$ emission times are given by \FIFRELIN using the process described in section \ref{sec:fifrelin}, with the CGCM level density model, and the EGLO model for the $\gamma$ transitions, to complete experimental data.}
	\label{fig:Ge_recoil_pred}
\end{figure}

Figure~\ref{fig:Ge_recoil_pred} shows that timing effects have an even greater impact on Ge cryogenic detectors since they induce three new prominent calibration features in addition to the 303~eV and 416~eV peaks of single-$\gamma$ transitions of $^{75}$Ge and $^{71}$Ge: two doublets of lines around 300~eV and 350~eV and one line at 561~eV (see Table~\ref{tab:calib_cascades}). This pattern offers a unique opportunity to study the response of the Ge detectors in the region of interest for low mass dark matter searches and detection of CE$\nu$NS with an accurate calibration of the cryogenic detector and a study of the evolution of the quenching factor in the sub-keV range when combining the heat and ionisation channels. However the separation of the different calibration features is quite demanding in terms of energy resolution. We discuss in section \ref{sec:discussion} how the detection of the high energy $\gamma$-rays in coincidence with the nuclear recoil could alleviate this constraint.

\begin{figure}
	\centering
 \includegraphics[width=1.0\linewidth]{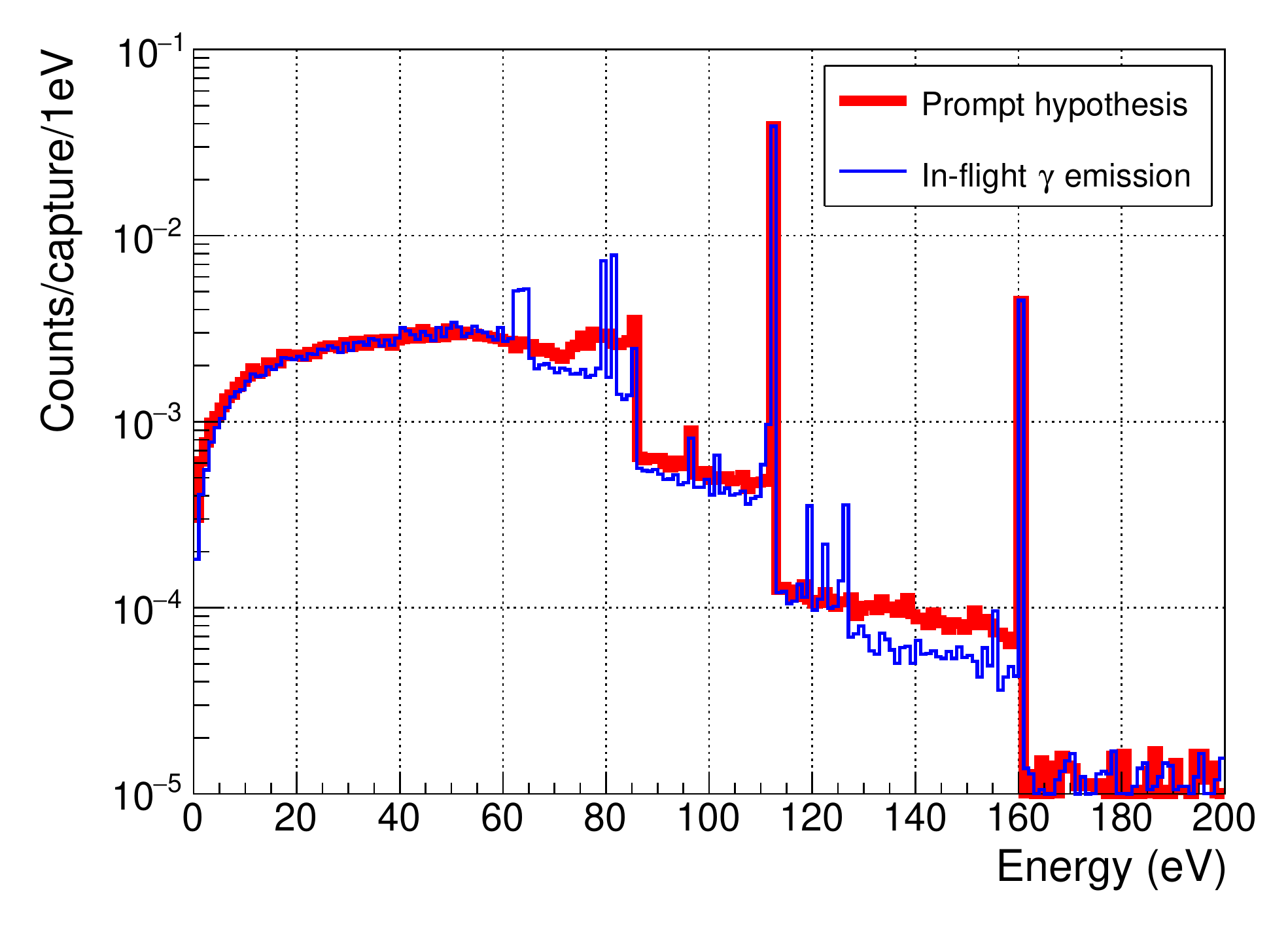}
	\caption{Predicted nuclear recoil spectra in a 5~$\times$~5~$\times$~5~mm$^3$ CaWO$_4$ crystal using the prompt hypothesis (red) or our best estimate of timing effects from the \FIFRADINA-\GEANT software (blue). The main impact of the timing effects is a more prominent peak structure around 80~eV, of crucial interest for linearity studies. This simulation was obtained with 2$\times$10$^7$ incident neutrons corresponding to 2.188$\times$10$^6$ neutron captures in the crystal. The $\gamma$ emission times are sampled by \FIFRELIN using the half-life times from the Weisskopf estimate to complete the experimental data.}
	\label{fig:W_recoil_pred}
\end{figure}

In the case of CaWO$_4$, used by the CRESST and NUCLEUS collaborations, the impact of timing effects is less important. However, we can see in Figure \ref{fig:W_recoil_pred} that the structure around 80~eV is much more prominent. This case is similar to the inset of figure \ref{fig:Al_recoil_pred}, a narrow doublet of lines (two 2-$\gamma$ cascades with a long-lived intermediate state) replaces the wider distribution predicted before \cite{Thulliez:2020esw}. 

\section{Discussion}
\label{sec:discussion}

We have shown that in the context of the CRAB calibration method the time dimension of the target nucleus recoil in the material as well as that of its $\gamma$-rays emission should be taken into account. We observe that the presence of de-excitation cascades with an energetic primary $\gamma$ and one or more low energy ($\lesssim$ 1 MeV) and long-lived (>few 10$^{-13}$~s) transitions is likely for most materials used in cryogenic detectors. In these configurations the target nucleus has time to stop before the next $\gamma$ emission which leads to a unique visible recoil energy for each cascade, independent of the $\gamma$ direction (Equation~\ref{eq:Ekin_sum}). When the cascade is probable enough this phenomenon provides a new calibration peak in the recoil spectrum and extends the potential of the \CRAB method.

\begin{figure}
	\centering
	\includegraphics[width=1.0\linewidth]{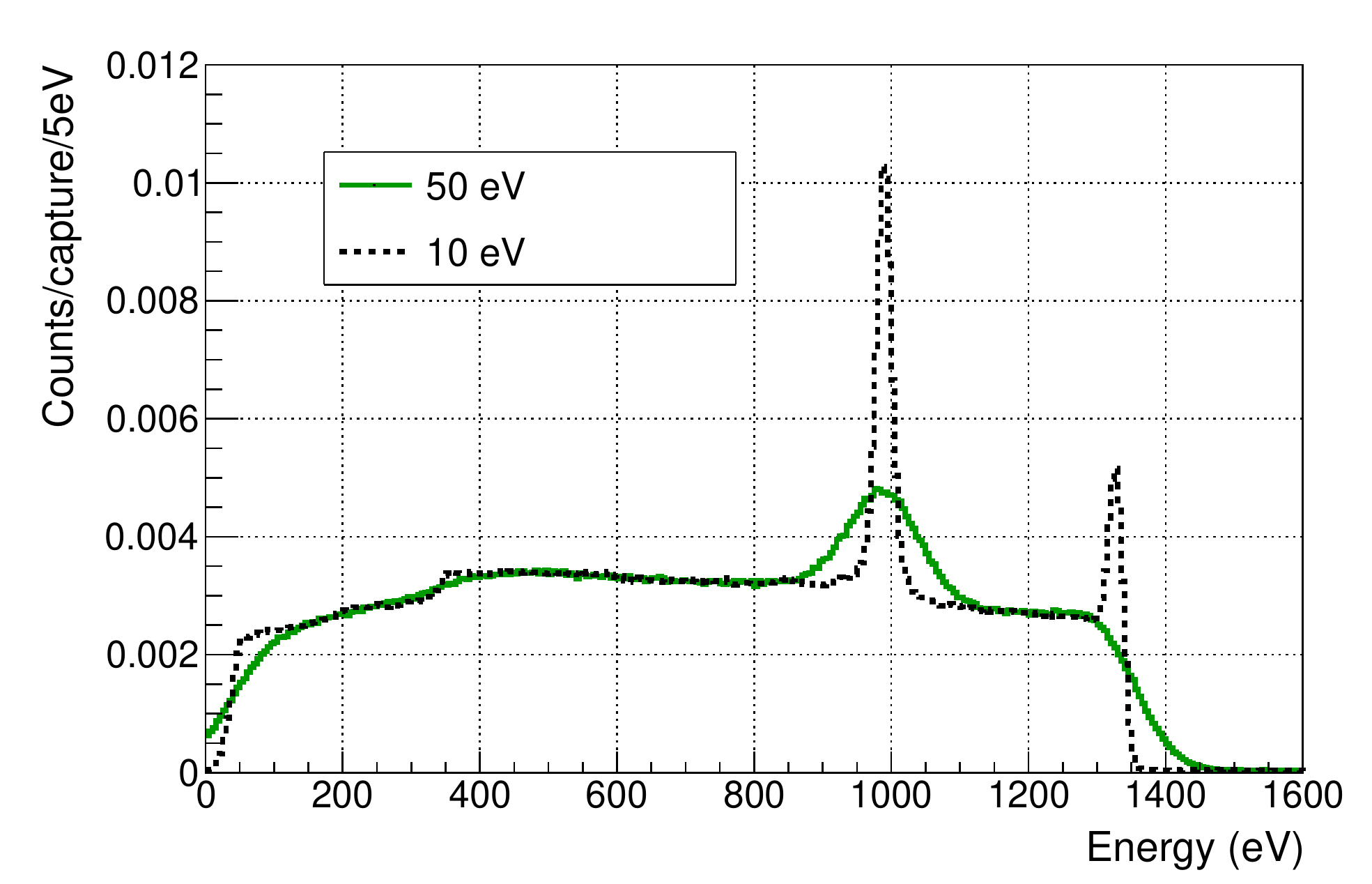}
	\caption{Predicted recoil spectrum in silicon with a constant energy resolution of 10~eV (dashed black) or 50~eV (plain green). The specific broader feature induced by the fine-tuned timing effects around the peak at 990~eV is far more robust than the single-$\gamma$-induced peak at 1330~eV with respect to energy resolution.}
	\label{fig:Si_50eV}
\end{figure}

In the case of Al$_2$O$_3$ crystals, the only significant new peak appears at an energy very close (9~eV) to the single-$\gamma$ peak and with a probability 10 times lower. It therefore provides little additional leverage for a study of the energy response of these detectors. The case of silicon is much more relevant with an additional peak at 990~eV, which is an intense and clearly separated from the single-$\gamma$ peak at 1330~eV. With a half-life of about 0.3 ps of the intermediate level of the 2-$\gamma$ cascade (see Table~\ref{tab:calib_cascades}), the emission of the second $\gamma$-ray occurs preferentially after the stop of the target nucleus or in flight at the end of its trajectory (see Figure~\ref{fig:Slowdown}). This results in a pedestal structure underneath the single-energy peak, which will be more resistant to energy resolution effects. Figure~\ref{fig:Si_50eV} shows that with a resolution of 50~eV the mono-energetic peaks are already no longer detectable while the pedestal structure remains clearly visible and allows an accurate calibration. Moreover the peak amplitude is expected to be sensitive to the slowing down process of the target nucleus in the material, which could provide a new test of the simulation codes. Using the coincident detection of the emitted primary $\gamma$ would even allow to determine the direction of the initial recoil and to study the dependence on this parameter.

\begin{figure}
	\centering
	\includegraphics[width=1.0\linewidth]{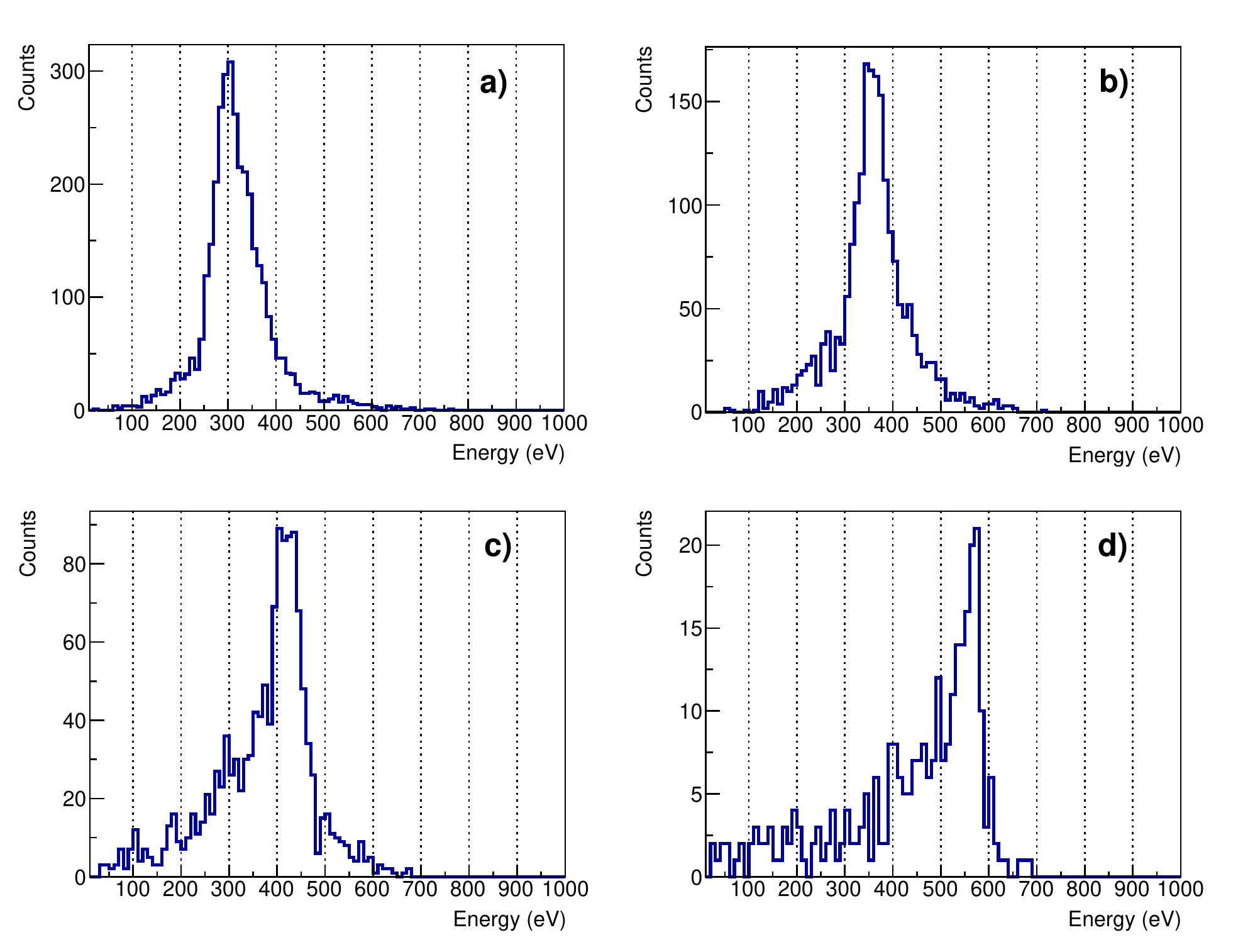}
	\caption{Predicted recoil spectra in a 20~$\times$~20~$\times$~20~mm$^3$ Ge crystal with a 20~eV constant energy resolution, when selecting the coincident detection of a primary $\gamma$ in a narrow energy window around the nominal energies shown in table \ref{tab:calib_cascades}: a) [6.0,6.5] MeV cut to select the 300.0~eV peak, b) [6.7,7.0] MeV cut to select the 353.1~eV peak, c) [7.3,7.6] MeV cut to select the 416.2~eV peak, d) [8.6,8.8] MeV cut to select the 561.2~eV peak. Clear peak structure appear for each selection allowing an accurate calibration at 4 different energies. In this simulation 10$^7$ neutrons were sent on the crystal, corresponding to 1.86$\times$10$^6$ captures and about 12 days of data taking with the experimental configuration proposed in \cite{Thulliez:2020esw}.}
	\label{fig:Ge_coinc_gamma}
\end{figure}

\begin{figure}
	\centering
	\includegraphics[width=1.0\linewidth]{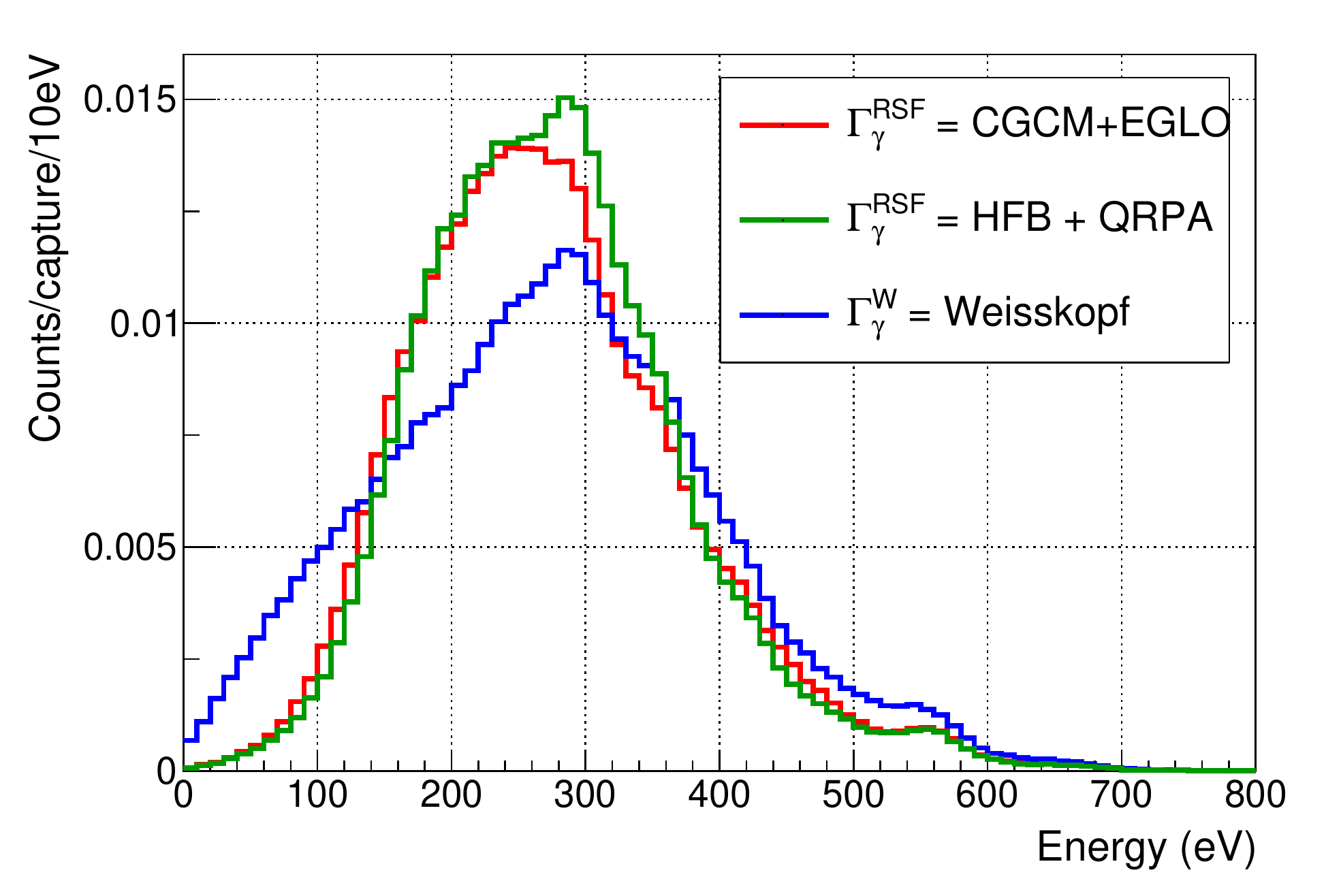}
	\caption{Predicted recoil spectra in a 20~$\times$~20~$\times$~20~mm$^3$ germanium crystal assuming a constant 20~eV resolution. Red: default \FIFRELIN configuration, the CGCM model is used for the nuclear level density and EGLO for the RSF; Green: the HFB model is used for the nuclear level density and QRPA for the RSF; Blue: Weisskopf estimate is used for all transitions with no experimental half-life.}
	\label{fig:Nuclear_Models}
\end{figure}

\noindent
Germanium, widely used in the cryogenic detector community, is probably the most promising with several new prominent peaks predicted in the recoil spectrum in the 300-600~eV range, relevant for the fundamental physics of light dark matter and CE$\nu$NS. However, the exploitation of the full richness of the recoil spectrum requires a very good energy resolution, typically 10~eV or better, which in practice would impose the use of small crystals. Exploiting the detection of the primary gamma in coincidence with the nuclear recoil significantly relaxes this constraint. Figure~\ref{fig:Ge_coinc_gamma} shows the result of a simulation where a thermal neutron beam is sent on a Ge crystal of 20~$\times$~20~$~\times$~20~mm$^3$ through the vessels of a dry cryostat. A gamma detector consisting in a simple assembly of 6~$\times$~6 PARIS phoswich detectors \cite{Ghosh:2016dtl} of total section 12"~$\times$~12" is placed on the floor under the cryostat. While with a 20~eV resolution Ge detector all calibration peaks are smeared out in the recoil spectrum (see Figure~\ref{fig:Nuclear_Models}), the selection of a detected $\gamma$ energy few~\% around the nominal energy of the primary $\gamma$ allows to recover the position of four calibration peaks with \%-level accuracy. Looking at Table~\ref{tab:calib_cascades} these peaks can be respectively identified to the first triplet of lines with weighted average recoil energy of 300.0~eV (taking into account few other sub-dominant lines in the same range), the lines doublet with weighted average of 351.1~eV and the two peaks corresponding to single-$\gamma$ transitions at 416.2~eV and 561.2~eV.
A detailed and accurate characterization of the detector response can be performed based on the position of these four peaks. A comparison between the phonon and ionization channels would also offer a unique study of the quenching factor at very low energy, still subject to large uncertainties \cite{Baxter:2022dkm}.

Without $\gamma$-tagging, Figure~\ref{fig:Nuclear_Models} shows that a 20 eV resolution already hinders the identification of calibration peaks in the recoil spectrum. One could think of using the edges and global maximum of the distribution to still get valuable information for calibration. However we show that the spectrum shape becomes quite sensitive to the nuclear models, as the continuum of high energy levels where the contribution of these models is large is also the main contributor to the continuous distribution of recoils under the calibration peaks. Thus the Ge recoil spectrum is particularly rich in underlying physics but we have the means to disentangle the different information. In the high energy resolution regime and/or with the coincident detection of the primary $\gamma$'s, an accurate study of the detector response can be performed since, as can be seen in Table~\ref{tab:calib_cascades}, most of the $\gamma$-transitions relevant for the Ge calibration peaks are well constrained experimentally and thus independent of the nuclear models. With a more modest resolution, or when focusing on the continuous contribution of recoils underneath the calibration peaks at high resolution, the global spectrum shape provides an original test of the nuclear models.

In principle, the dependence of the shape of the continuous spectrum of CaWO$_4$ on nuclear models should be similar to that of Ge. For the time being, priority has been given to the Ge case for these time consuming computations with different nuclear models, and the CaWO$_4$ has been treated with the Weisskopf approximation only. However calculations with RSF nuclear models could be applied to the four relevant tungsten isotopes as well for the analysis of future data. Within the Weisskopf approximation the predicted enhancement of the 80~eV peak due to timing effects is already a valuable result. It will extend the range of calibration of CaWO$_4$ detectors and allow an accurate study of the linearity at unprecedented low energy. The sensitivity to the creation of a single crystal defect could be within reach with unique tests of the underlying solid state physics.

\section{Conclusion}

In summary, we have developed simulations coupling the \FIFRELIN, \IRADINA and \GEANT codes to provide reference predictions of the nuclear recoils induced by neutron capture in cryogenic detectors. We have shown that a detailed description of the development in time of the de-excitation $\gamma$-cascade and the atomic displacements in matter is mandatory for accurate prediction of the spectra shape. While the positions of the calibration peaks discussed in our previous article \cite{Thulliez:2020esw} are not affected by this improvement, new peaks and spectrum features in general emerge from these timing effects. Thus we expect very rich physics from the measurement of recoil spectra induced by neutron capture in cryogenic detectors in the coming years. Different detection techniques can be used to disentangle all effects: with state-of-the-art energy resolution or by tagging the primary de-excitation $\gamma$-ray in coincidence with the nuclear recoil, the determination of the position of the recoil peaks provide an accurate calibration of the cryogenic detectors with direct application to the CE$\nu$NS and light dark matter searches. The physics case for the germanium based detectors is particularly interesting with a potentially unique study of quenching factors in the sub-keV range. The regime of the lowest energies of nuclear recoils in CaWO$_4$ crystals could provide an original probe of solid state physics and associated simulations. Finally we have shown with the detailed study of the Ge spectra that the continuous distribution of nuclear recoils under the calibration features, remaining accessible even with a more modest energy resolution, has a unique sensitivity to the nuclear models of level densities and radiative strength functions.

\textbf{Data Availability Statement} This manuscript has associated data in a data repository. We make available millions of de-excitation cascades with associated nuclear recoil energies for each isotope at \url{https://doi.org/10.5281/zenodo.7936552}, since other running and upcoming projects might profit from these data as well.

\bibliographystyle{unsrtnat}
\bibliography{./timing_crab_paper}

\end{document}